\nonstopmode
\documentclass[amsfonts,showpacs,tightenlines,aps,12pt]{revtex4}
\usepackage{bm}

\newcommand{\bea}{\begin{eqnarray}}
\newcommand{\eea}{\end{eqnarray}}
\newcommand{\nn}{\nonumber} 
\newcommand{\rhov}{\bm{\rho}}  
\begin{document}

%\preprint{TRIUMF preprint TRI-PP-00-49, submitted to Phys.~Rev.~C}

\title{Spectroscopic amplitudes and microscopic substructure effects in
nucleon capture reactions}
\author{Jutta Escher}\email{escher@triumf.ca}
\author{Byron K.~Jennings}\email{jennings@triumf.ca}
\affiliation{TRIUMF, 4004 Wesbrook Mall, Vancouver, BC, Canada V6T 2A3}
\author{Helmy S.~Sherif}\email{sherif@phys.ualberta.ca}
\affiliation{TRIUMF, 4004 Wesbrook Mall, Vancouver, BC, Canada V6T 2A3 \\ 
\& Department of Physics, University of Alberta, Edmonton, Alberta,
Canada, T6G 2J1 
\footnote{permanent address}  }
\date{\today}

\begin{abstract}
Spectroscopic amplitudes play an important role in nuclear capture reactions.
These amplitudes are shown to include both single-particle and polarization
effects: the former through their spatial dependence and the latter through
their normalization (the spectroscopic factors).  Coupled-channels equations
are developed for the spectroscopic amplitudes.  These equations serve as a
convenient starting point for the derivation of several approximations:
Hartree, Hartree-Fock and two different single-particle models.  The
single-particle models include antisymmetry in different ways, but both miss
many-body effects.  Therefore, cross sections calculated with either of these
models need to be multiplied by the spectroscopic factor.
\end{abstract}


\maketitle

\section{Introduction}

Nucleon capture reactions at low energies, such as $^7$Be$(p,\gamma)^8$B,
$^{16}$O$(p,\gamma)^{17}$F$^*$, or $^7$Li$ (n,\gamma)^8$Li, play an important
role in our understanding of astrophysical phenomena.  For example, in the
hydrogen-burning process in stars such as our sun, low-energy proton capture on
beryllium takes place in the proton-proton chain, and the
$^{16}$O$(p,\gamma)^{17}$F$^*$ reaction occurs in the CNO
cycle~\cite{AstroBkgrnd,Adelberger98,Morlock97}.  Exact knowledge of the
reaction rates is necessary for modeling the energy generation and evolution of
hydrogen-burning stars.  In addition, the $^7$Be$(p,\gamma)^8$B reaction at
solar energies $(E_{cm}\leq 20$ keV) plays a key role in the `solar neutrino
puzzle' since the neutrino event rate in the existing chlorine and water
Cerenkov detectors is dominated by the high-energy neutrinos produced in the
subsequent $\beta$ decay of $^8$B~\cite{AstroBkgrnd,Adelberger98,Bahcall89}.
The $^7$Li$ (n,\gamma)^8$Li reaction is a key element of primordial
nucleosynthesis in inhomogeneous big bang scenarios~\cite{AstroBkgrnd,BigBang}.
It initiates a sequence of reactions which bridge the mass gap at $A$=8 and
thus its rate is crucial for determining the amounts of heavier elements
produced in these models.

Direct measurements of capture reactions at energies corresponding to
astrophysically relevant temperatures, however, are often very difficult, since
the cross sections diminish exponentially at low energies.  Thus, theoretical
studies of these processes become very valuable.  In addition to microscopic
theories, such as the nuclear shell model or cluster models, one-body potential
models provide a popular framework for such investigations.  For example, the
potential model was used in ref.~\cite{Jenn98} to discuss the energy dependence
of the reaction rates.  In the one-body potential model, a single-particle wave
function is used to calculate various observables; microscopic substructure
effects are partially accounted for through the use of spectroscopic factors.
This strategy, however, has recently been the subject of vivid
discussions~\cite{Csoto00,Muk01}.  At issue is the proper normalization of the
cross section.  Cs\'{o}t\'{o}~\cite{Csoto00}, for instance, maintains that
spectroscopic factors should not be included in potential-model calculations of
the $^7$Be$(p,\gamma)^8$B cross section at low energies, since the reaction
depends only on the asymptotic normalization coefficient of the $^8$B
bound-state wave function~\cite{Muk99}, whereas the spectroscopic factor arises
from the short-range properties of the wave function.  Mukhamedzhanov {\em et
al.}~\cite{Muk01,Muk99} argue in favor of using a different approach, based on
asymptotic normalization coefficients instead of spectroscopic factors, for
determining the relevant cross sections.  However, the asymptotic normalization
coefficient contains short-range effects.  It can actually be given in terms of
an integral over the interior of the nucleus~\cite{Jennings00} and its
interdependence with the spectroscopic factor has been noted in the earlier
work of Locher and Mizutani~\cite{Locher} and of Lovas {\em et
al.}~\cite{Lovas87}.

These recent discussions have motivated us to revisit the question of the
proper treatment of microscopic nuclear structure effects in one-body models.
In the present work we focus on the role of spectroscopic amplitudes and
factors.  The use of spectroscopic factors in nuclear reaction calculations
dates back to the early days of nucleon transfer reactions~\cite{Austern} and
continues to be central in the interpretation of such processes~\cite{eep}.
With the renewed interest in nucleon capture reactions in the context of
astrophysical scenarios, and the emerging need for very accurate reaction
rates, it becomes necessary to review and clarify the assumptions associated
with one of the most basic models of nuclear physics, the one-body potential
model.

Before proceeding with the formalism, it is necessary to clarify the meaning of
the terms spectroscopic amplitude, spectroscopic factor, and one-body model.
For example, it is important to realize that there are many spectroscopic
amplitudes and spectroscopic factors associated with an $A$-body system, namely
one corresponding to each excited state of the $(A-1)$-body system.  The
different amplitudes are not independent since they are related by a
model-independent sum rule~\cite{Die}.  In addition, the spectroscopic
amplitudes are related by a set of coupled-channels equations.

The question of what is meant by a one-body model is more complicated since
different notions are associated with the term.  The coupled-channels formalism
presented here is used to derive two different one-body models and to study the
connections between them.  One approach, which uses either the Hartree or the
Hartree-Fock approximation in an intermediate step, leads to $A$
noninteracting nucleons in a one-body potential.  This, strictly speaking, is
not a one-body model since there are still $A$ particles.  However, the
orbitals corresponding to the different particles decouple and the problem
reduces to solving $A$ one-body equations.  The fact that we still have $A$
nucleons explicitly present has the advantage that antisymmetry can be built in
{\em ab initio} using a Slater determinant for the $A$-body wave function.
Another method for generating a one-body model is based on integrating out the
coordinates of $(A-1)$ nucleons, effectively projecting onto a low dimensional
space.  In this approach antisymmetry is enforced separately for each channel
in the coupled-channels equations.  Truncating to a single channel results in
the approximation advocated by Varga and Lovas {\em et al.}~\cite{Lovas}, 
who study the cluster substructure of $^6$Li in this framework.  
Their work is based on the generator coordinate method.  
The derivation presented here serves to clarify
the relation between their model and other approximations.

In Section~\ref{sec:sa}, we define the terms spectroscopic amplitude and
spectroscopic factor, derive coupled-channels equations for the amplitudes, and
discuss several approximation schemes, including the first one-body model
described above.  An alternative approach to the spectroscopic amplitudes is
considered in Section~\ref{sec:aa}, which includes details on the second
one-body model mentioned above.  In Section~\ref{sec:Csoto}, we explore the
physical aspects associated with the spatial dependence of the spectroscopic
amplitude and its norm, the spectroscopic factor.  We give an expression for
the reaction matrix elements in terms of the spectroscopic amplitudes.  For
both single-particle models considered here the reactions rates need to be
multiplied by a spectroscopic factor.  Our conclusions are summed up in
Section~\ref{sec:Concl}, and various technical aspects of our work are included
in the Appendices.

\section{Spectroscopic Amplitudes}
\label{sec:sa}

In this section we expand an $A$-body wave function in terms of the ground and
excited states of the ($A$-1)-body system and derive a set of coupled
differential equations for the expansion coefficients, of which the
spectroscopic amplitudes are a special case.  The resulting equations of motion
give insight into the long-range behavior of the $A$-body ground state wave
function and lead to various approximation schemes.  In particular, the Hartree
and Hartree-Fock approximations and a one-body potential model approach emerge
naturally.  For simplicity, spin and isospin degrees of freedom are suppressed
and the Coulomb potential is neglected.  Furthermore, we do not consider the 
center-of-mass motion here; this subject is discussed in 
Appendix~\ref{App:COMcorr}.

Let $\Psi^n_{A-1} (\bm{r}_1,\cdots,\bm{r}_{A-1})$ denote the $n$-th (fully
antisymmetric) eigenstate of the $(A-1)$-body Hamiltonian $H_{A-1}$.  The
collection of all $\Psi^n_{A-1}$, including both bound and continuum states,
forms a complete set of states in the space of antisymmetric $(A-1)$-body wave
functions.  Using this set, $\{\Psi^n_{A-1} \}_{n=1,2,\ldots}$, one can
construct a basis for the space of antisymmetric $A$-body wave functions, by
defining $_{\cal A}\Psi^{n,r}_A (\bm{r}_1,\cdots,\bm{r}_A)=$ ${\cal
A}\Psi^{n,r}_A(\bm{r}_1,\cdots,\bm{r}_A)$, where $\Psi^{n,r}_A(\bm{r}_1,
\cdots,\bm{r}_A) \equiv \Psi^n_{A-1}(\bm{r}_1,\cdots,\bm{r}_{A-1})
\delta(\bm{r}-\bm{r}_A)$ and ${\cal A}$ antisymmetrizes between the $A$-th
coordinate, which occurs in the delta function, and the $(A-1)$ coordinates in
$\Psi^n_{A-1}$.  This `intercluster' antisymmetrization operator is normalized
so as to satisfy ${\cal A}^2=\sqrt{A}{\cal A}$.

The $\Psi^{n,r}_A$ span a space that includes both totally antisymmetric
$A$-body states and mixed-symmetric states, which are antisymmetric in the
first $A$-1 coordinates and symmetric under interchanges involving the $A$-th
coordinate.  The basis sets $\{ \Psi^n_{A-1} \}$, $\{ \Psi^{n,r}_A
\}$, and $\{_{\cal A}\Psi^{n,r}_A \}$ are discussed in more
detail in Appendix~\ref{sec:App_bases}.

An arbitrary antisymmetric $A$-body wave function $\psi_A$ can then be expanded
as:
\bea
|\psi_A\rangle &=& \frac{1}{A} \sum_{n=1}^\infty \int d\bm{r} |_{\cal
A}\Psi^{n,r}_A\rangle \phi_n(\bm{r}) \nn \\ 
               &=& \frac{1}{\sqrt{A}} \sum_{n=1}^\infty \int d\bm{r}
        |\Psi^{n,r}_A\rangle \phi_n(\bm{r})
\label{eq:expn} \\           
               &=& \frac{1}{\sqrt{A}} \sum_{n=1}^\infty |\Psi^n_{A-1}\rangle
|\phi_n\rangle  
\nn 
\eea 
with expansion coefficients:
\bea
\phi_n(r) &=& \langle_{\cal
A}\Psi^{n,r}_A|\psi_A\rangle
\nn \\
            &=&
\sqrt{A}\langle\Psi^{n,r}_A|\psi_A\rangle 
\\
            &=& \langle\Psi^n_{A-1}| a(\bm{r})
|\psi_A\rangle \; .
\nn
\eea
Here $a(\bm{r})$ [$a^{\dagger}(\bm{r})$] is an annihilation [creation] operator
which destroys [creates] a nucleon at position $\bm{r}$ and obeys the usual
anticommutation relations.  Note that only the first expansion in
eq.~(\ref{eq:expn}) is manifestly antisymmetric, while in the other two
expressions the antisymmetry information is carried by the expansion
coefficients.  The coefficients are identical in all three expansions if and
only if $\psi_A$ is fully antisymmetric.  In this case, the mixed-symmetry
components of $\Psi^{n,r}_A$ do not contribute.

When $\psi_A$ denotes a bound state, the $\phi_n(\bm{r})$ are called {\em
spectroscopic amplitudes}, and the associated integrals
\bea
S_n &=& \int d\bm{r} \left|\phi_n(\bm{r})\right|^2
\eea
are the frequently used {\em spectroscopic factors}~\cite{B+G77}.  They obey the
sum rule $\sum_{n=1}^\infty S_n=A$ (see ref.~\cite{Die} and also
eq.~(\ref{eq:pnorm})).  We observe that for a bound $A$-body state, there are
many spectroscopic amplitudes (and thus many spectroscopic factors), namely one
for each excited state of the $(A-1)$-body nucleus.  Given the structural
information on the $(A-1)$-body system that enters the wave functions
$\Psi^n_{A-1}$, the spectroscopic amplitudes completely determine the wave
function $\psi_A$.  The spatial dependence of $\phi_n(\bm{r}_A)$ is related to
the properties of the single-particle orbital of the $A$-th nucleon in the
larger system, and the norm of $\phi_n(\bm{r})$, the spectroscopic factor,
provides a measure of the structural similarity of the $n$-th excited
$(A-1)$-body state and an $(A-1)$-body subcluster of the full $A$-nucleon
system.

The interdependence of the spectroscopic amplitudes can be made explicit upon
deriving a set of differential equations for the $\phi_n(\bm{r})$.  For an
$A$-body Hamiltonian $H_A$, which contains a kinetic energy term and a two-body
potential, $H_A=-\sum_{i=1}^{A}\frac{\nabla^2_{r_i}}{2m_i} + \frac{1}{2}
\sum_{i,j=1}^{A} V(|\bm{r}_i - \bm{r}_j|)$, we can write
$H_A = H_{A-1}-\frac{\nabla^2_{r_A}}{2m_A}+\sum_{i=1}^{A-1} V(|\bm{r}_i -
\bm{r}_A|)$, where $H_{A-1}$ denotes the Hamiltonian of the $(A-1)$ body system
and $m_i$ is the mass of the $i$-th nucleon (for simplicity we assume $m_i=m$
for $i=1,\ldots,A$).  Since $|\Psi^n_{A-1}\rangle$ and $|\psi_A\rangle$ are
eigenstates of $H_{A-1}$ and $H_A$, respectively, we have:
\bea 
\langle\Psi^n_{A-1}|H_A| \psi_A \rangle&=& 
          E_A\langle\Psi^n_{A-1}|\psi_A\rangle
          \nonumber \\ &=&
E_{A-1}^n\langle\Psi^n_{A-1}|\psi_A\rangle +
\langle\Psi^n_{A-1}|-\frac{\nabla^2_{r_A}}{2m}+\sum_{i=1}^{A-1}
          V(|\bm{r}_i - \bm{r}_A|)|\psi_A\rangle \; .
\eea
Inserting the expansion given in eq.~(\ref{eq:expn}) for $|\psi_A\rangle$, we
obtain a set of exact, Schr\"odinger-like, coupled equations for the
spectroscopic amplitudes:
\bea
(E_A-E_{A-1}^n)\phi_n(\bm{r}) =
-\frac{\nabla^2_{r}}{2m} \phi_n(\bm{r}) 
       + \sum_{n'=1}^\infty \langle\Psi^n_{A-1}|
          \sum_{i=1}^{A-1} V(|\bm{r}_i - \bm{r}|)|
            \Psi^{n'}_{A-1}\rangle \phi_{n'}(\bm{r})  .
\label{eq:schro}
\eea
This set of equations, originally derived for stripping reactions (see for
example ref.~\cite{pinkston}), is not sufficiently appreciated. In the form
just given, these equations may be too complex for use directly in
calculations, since the $V_{nn'}(\bm{r}) \equiv \langle\Psi^n_{A-1} |
\sum_{i=1}^{A-1} V(|\bm{r}_i - \bm{r}|)| \Psi^{n'}_{A-1}\rangle$ couple an
infinite set of coefficients.  However, they allow us to discuss the
long-range behavior of the $A$-body ground state wave function as well as
derive commonly used approximations. 

The solutions of eq.~(\ref{eq:schro}) include not only the physically relevant
completely antisymmetric states, but also the mixed-symmetric, i.e., unphysical,
states.  As long as the coupled equations are solved exactly this does not
cause any problems since the two classes of states do not mix.  However, when
approximations are invoked one needs to ensure that the unphysical solutions
are eliminated.

From Eq.~(\ref{eq:schro}) we can extract information on the long-range behavior
of the $A$-body ground state wave function.  The matrix element
$V_{nn'}(\bm{r}_A)$, for $n$ or $n'$ corresponding to an $(A-1)$-body bound
state, has a range which is determined by the convolution of the (short-range)
two-body potential with the bound-state wave function.  Thus it falls off
rapidly with increasing $r_A=|\bm{r}_A|$.  When both $\Psi^n_{A-1}$ and
$\Psi^{n'}_{A-1}$ describe continuum states, the matrix element has a 
long-range, but infinitesimal, tail.  
Solving eq.~(\ref{eq:schro}) outside the range of the
potential, we find that the $\phi_n(\bm{r}_A)$ decouple and fall off
exponentially, $\phi_n(\bm{r}_A) \propto \exp[-\sqrt{2 m (E^n_{A-1}-E_A)} \,
r_A]/r$.  Since $(E^n_{A-1}-E_A)$ is smallest for $n=1$, the long-range
behavior of the A-body wave function is dominated by $\phi_1
(\bm{r}_A)\Psi^1_{A-1} (\bm{r}_1,\cdots,\bm{r}_{A-1})$.  Thus processes with
reaction probabilities that are peaked in the asymptotic region depend only on
the asymptotic normalization coefficient, $A_{nc}=\lim_{r\rightarrow\infty}
\phi_1(\bm{r}) r \exp[\sqrt{2 m (E_A-E_{A-1}^1) r}]$ (see also
refs.~\cite{Muk01,Muk99}). 
The asymptotic normalization coefficient is a property of the spectroscopic 
amplitude and hence implicitly includes the spectroscopic factor.

To obtain an explicit expression for the asymptotic normalization coefficient,
we Fourier transform eq.~(\ref{eq:schro}) (to include center-of-mass
corrections use eq.~(\ref{eq:intschro})). This gives us:
\bea
\tilde\phi_n(\bm{k})=\frac{1}{E_A-E_{A-1}^n-\frac{k^2}{2m}}
\sum_{n'=1}^\infty
\int  d\bm{r} \; e^{i\bm{k}\bm{r}}
\langle\Psi^n_{A-1}| 
          \sum_{i=1}^{A-1} V(|\bm{r}_i - \bm{r}|)| \Psi^{n'}_{A-1}\rangle
            \phi_{n'}(\bm{r}) .
\eea
The pole in $\tilde\phi_n(\bm{k})$ is thus explicitly seen. It is this pole
which is responsible for the upturn in the astrophysical $S$ factor of the
$^7$Be$(p,\gamma)^8$B and $^{16}$O$(p,\gamma)^{17}$F$^*$ reactions as the
incident momentum goes to zero~\cite{Jenn98}.  As shown in
ref.~\cite{Jennings00}, the residue at this pole for $n=1$ is proportional to
the asymptotic normalization coefficient. We have:
\bea
A_{nc}=\lim_{k\rightarrow -i \kappa}4m\pi^2
\sum_{n'=1}^\infty
\int  d\bm{r} \; e^{i\bm{k}\bm{r}}
\langle\Psi^1_{A-1}| 
          \sum_{i=1}^{A-1} V(|\bm{r}_i - \bm{r}|)|
            \Psi^{n'}_{A-1}\rangle \phi_{n'}(\bm{r})  \; ,
\eea
where $\kappa=\sqrt{2m(E_A-E_{A-1}^1)}$, i.e. the asymptotic normalization
coefficient can be given as an integral over the interior of the nucleus
(since $\Psi^1_{A-1}$ has the spatial extent of the ground state of the
$(A-1)$-body system).  The price paid is that there is 
a sum over all the spectroscopic amplitudes.

For arbitrary distances, we may consider approximating the $\Psi^n_{A-1}$ in
eq.~(\ref{eq:schro}) by Slater determinants constructed from the spectroscopic
amplitudes $\phi_{i}(\bm{r})$ with $i \in \{1,\ldots,A \}$.  In this case, we
obtain the Hartree-Fock equations in coordinate space.  The $A$ Hartree-Fock
orbitals can thus be identified as approximations to the spectroscopic
amplitudes and the Hartree-Fock single-particle energies approximate
single-nucleon separation energies.  The local (Hartree) term arises from the
diagonal terms ($V_{nn}$) in eq.~(\ref{eq:schro}), and the off-diagonal terms
($V_{nn'}$ with $n \neq n'$) give the nonlocal (exchange) potential, i.e., the
Fock terms originate from the channel coupling.  If the off-diagonal elements
($V_{nn'}$, $n \neq n'$) are neglected we obtain the Hartree approximation.

A one-body potential model, which treats the nucleus as a system of $A$ {\em
noninteracting} nucleons, can be obtained from either the Hartree or the
Hartree-Fock approximation.  Starting from the Hartree equations, one can
express the many-body wave functions $\Psi^n_{A-1}$ as Slater determinants of
the $\phi_{i}(\bm{r})$, $i \in \{1,\ldots,A \}$, and impose the condition that
the one-body potential be the same for each single-particle orbital,
$V_{nn}(\bm{r}) \Rightarrow V(\bm{r})$.  Alternatively, beginning from the
Hartree-Fock equations, one can dictate the replacement:
\bea
V_{nn'}(\bm{r}) = \langle\Psi^n_{A-1}|\sum_{i=1}^{A-1} V(|\bm{r}_i -
\bm{r}|)|\Psi^{n'}_{A-1}\rangle \Rightarrow
V(\bm{r})\delta_{nn'} \; .
\eea
In both cases, one obtains the following equation for the $\phi_n(\bm{r})$:
\bea
(E_A-E_{A-1}^n)\phi_n(\bm{r}) &=& -\frac{\nabla^2_{r}}{2m}\phi_n(\bm{r}) +
V(\bm{r})\phi_{n}(\bm{r}) \; ,
\label{eq:1bSchro} 
\eea  
which defines a one-body potential model with $A$ non-interacting nucleons
in a common potential.  Thus, the one-body potential model may be regarded as
an approximation to either the Hartree or the Hartree-Fock approach.  In this
model, antisymmetry can be ensured in a straightforward manner: many-body
wave functions, such as $\Psi^n_{A-1}$ or $\psi_A$, are -- like in the
Hartree-Fock picture -- Slater determinants constructed from $A$ given 
single-particle wave functions.

The Slater determinants, in one-body potential models, usually play only a
formal role and disappear from sight in actual calculations, to the extent that
their existence is often forgotten causing conceptual confusion. For the matrix
elements of a one-body operator, $ {\cal O}({\bm r})$, between states that
differ only in one orbital, the remaining $(A-1)$-body orbitals integrate out
leaving just the active orbitals. Thus we have a matrix element like $\int
d{\bm r} \phi_m({\bm r}) {\cal O}({\bm r})\phi_{m'}({\bm r})$ where $m$ and $m'$
denote the active orbitals. At first sight this expression is a pure one-body
expression that seems to have no antisymmetry present and indeed at the
computational level this is true. In calculating the matrix elements of
one-body operators there is no need to explicitly consider the Slater
determinants. However, the Slater determinants did their job of ensuring
antisymmetry before they were integrated out.

Since the single-particle Schr\"odinger equation, eq.~(\ref{eq:1bSchro}),
originates from the coupled-channels equations for the expansion coefficients,
eq.~(\ref{eq:schro}), the bound-state single-particle wave functions of the
potential model discussed here approximate the spectroscopic amplitudes.
Therefore, the potential-model approximation gives $A$ nonzero spectroscopic
amplitudes, all of which are normalized to one. Consequently, the associated
spectroscopic factors are one and carry no information on the many-body
correlations of the nuclei involved.

One can, however, move beyond the simple picture of $A$ noninteracting
particles in a common potential and include some many-nucleon correlations in a
schematic manner.  As a first correction, one usually allows for
single-particle orbitals which are different for the $(A-1)$-body and $A$-body
systems.  As a result, the expansion of $\psi_A$, eq.~(\ref{eq:expn}), will
have contributions from more than $A$ terms.  Consequently, the spectroscopic
factors will be, on average, less than one, and most individual factors will be
smaller than one.

Another important correction to be taken into account involves the treatment of
the center-of-mass motion and leads to the introduction of an intrinsic
spectroscopic amplitude (see Appendix~\ref{App:COMcorr}).  In both the
Hartree-Fock and the one-body potential-model approaches, the center of mass is
erroneously confined in a potential.  In potential models which are based on a
harmonic oscillator potential the center-of-mass corrections can be taken into
account exactly.  In this case, the spectroscopic amplitudes corresponding to
valence shell orbitals increase by a factor $[A/(A-1)]^{m/2}$, where $m$
denotes the major oscillator shell under consideration, while the amplitudes
associated with lower shells decrease accordingly~\cite{Die}.  The sum rule for
the spectroscopic factors remains unaffected. The center-of-mass correction
causes some of the spectroscopic factors to be greater than unity.

The above considerations illustrate that information on the structure of the
$A$-body wave function of a nuclear system, which is lost in a simple one-body
potential model approach, can be in part recovered through the appropriate
correction to the single particle wave functions, that is by multiplying by the
spectroscopic factors.  In Section~\ref{sec:Csoto} we will discuss how the
resulting deviations of the spectroscopic factors from one (or zero) affect
physical observables such as the cross sections for external capture reactions.

\section{An Alternative Perspective}
\label{sec:aa}

In this section we approach the spectroscopic amplitudes from a different
perspective.  Motivated by cluster-model results, we derive a set of coupled
integral equations.  While general cluster-model calculations involve two or
more clusters of arbitrary size, we restrict ourselves to describing an
$A$-body nucleus as an ($A$-1)-body cluster plus a single nucleon.  The
approach pursued in this section has the advantage that explicit channel
coupling is no longer required to ensure antisymmetry.  Again, various
approximation schemes can be used to simplify the coupled equations.  In
particular, in one of these schemes the Hartree-Fock equations can be
recovered; in another a one-body model emerges, which was previously discussed
by various authors~\cite{Lovas,Saito} and which differs from the model
presented in the last section.

Starting with the Schr\"{o}dinger equation for an $A$-body system, 
$H_A\psi_A=E_A\psi_A$, and expanding $\psi_A$ in terms of the antisymmetric
basis states $_{\cal A}\Psi^{n,r}_A$, we obtain:
\bea
\sum_{n'=1}^\infty\int d\bm{r}'
\langle_{\cal A}\Psi^{n,r}_A|H_A|_{\cal A}\Psi^{n',r'}_A\rangle\phi_{n'}(\bm{r}')
&=& E_A  \sum_{n'=1}^\infty\int d\bm{r}' 
\langle_{\cal A}\Psi^{n,r}_A| _{\cal A}\Psi^{n',r'}_A\rangle\phi_{n'}(\bm{r}') \; ,
\label{eq:schro1a}
\eea
or, equivalently:
\bea
\sum_{n'=1}^\infty\int d\bm{r}'{\cal _AH}(n,\bm{r},n',\bm{r}')\phi_{n'}(\bm{r}') 
&=& E_A \sum_{n'=1}^\infty\int d\bm{r}' {\cal N}(n,\bm{r},n',\bm{r}')
\phi_{n'}(\bm{r}') \; ,
\label{eq:schro1b}
\eea
where 
${\cal _AH}(n,\bm{r},n',\bm{r}')=\langle_{\cal A}\Psi^{n,r}_A|H_A|_{\cal A}
\Psi^{n',r'}_A\rangle$ and 
${\cal N}(n,\bm{r},n',\bm{r}')=\langle_{\cal A} \Psi^{n,r}_A| _{\cal A}
\Psi^{n',r'}_A\rangle$ denote the kernels of integral operators 
${\cal _A}\hat{{\cal H}}_{nn'}$ and $\hat{{\cal N}}_{nn'}$, respectively. 
The operator $\hat{{\cal I}}_{nn'}$ ($\hat{{\cal I}}_{nn'}=$ 
${\cal _A}\hat{{\cal H}}_{nn'}$ or $\hat{{\cal N}}_{nn'}$), which is 
the $(n,n')$ entry of an infinite-dimensional matrix $\hat{{\cal I}}$
($\hat{{\cal I}}=$ ${\cal _A}\hat{{\cal H}}$ or $\hat{{\cal N}}$), acts as 
follows:  $\hat{{\cal I}}_{nn'} f(\bm{r}) \equiv \int d\bm{r}' {\cal I}
(n,\bm{r},n',\bm{r}') f(\bm{r}')$.   The matrix $\hat{{\cal N}}$ serves as
the norm operator for the antisymmetrized basis $\{ _{\cal A}\Psi^{n,r}_A \}$
and has many interesting properties which are discussed in
Appendix~\ref{sec:App_normOp}.  $\hat{{\cal N}}/A$ is a
projection operator which projects an arbitrary many-body state into the set of
completely antisymmetric wave functions.  Thus, eqs.~(\ref{eq:schro1a}) and
(\ref{eq:schro1b}) contain an explicit projection onto completely antisymmetric
states, in contrast to the coupled differential equations discussed in the
previous section, eq.~(\ref{eq:schro}).  This projection means that the
solutions of eq.~(\ref{eq:schro1b}), $\phi_{n}^{(sol)}(\bm{r})$, can contain 
arbitrary contributions from mixed-symmetric states while still satisfying the 
equations of motion. Thus the projected state, 
$\phi_{n} = (1/A) \sum_{n'=1}^\infty\int d\bm{r}' {\cal N} (n,\bm{r},n',\bm{r}')
\phi_{n'}^{(sol)}(\bm{r}')$, rather than $\phi_{n}^{(sol)}(\bm{r})$ corresponds 
to the physical state.

With eq.~(\ref{eq:schro1a}) (or, equivalently, eq.~(\ref{eq:schro1b})), we have
derived a set of coupled-channels (integral) equations, analogous to the system
of coupled differential equations in the previous section,
eq.~(\ref{eq:schro}).  In the present approach, however, antisymmetry is
explicitly enforced, resulting in a set of equations which contain a projection
operator and a complicated expression for the effective Hamiltonian.  To
illustrate this, we compare the Hamiltonian kernel in the nonantisymmetrized
basis, ${\cal H}(n,\bm{r},n',\bm{r}')$, with the corresponding expression in
the antisymmetrized basis, ${\cal _AH}(n,\bm{r},n',\bm{r}')$.  The former is
local, i.e., diagonal, in the spatial coordinate, although not in the discrete
variable $n$,
\bea
{\cal H}(n,\bm{r},n',\bm{r}')&=&\langle\Psi^{n,r}_A|H_A|\Psi^{n',r'}_A
\rangle \\ 
&=& \delta(\bm{r}-\bm{r}')\left[ \delta_{nn'} 
\left( E_{A-1}^n  - \frac{\nabla^2_{\bm{r}'}}{2m} \right) 
+ \langle \Psi^n_{A-1}| \sum_{i=1}^{A-1} V(|\bm{r}_i - \bm{r}|)
| \Psi^{n'}_{A-1} \rangle \right]  \, ,
\label{eq:ham}
\eea 
whereas the latter has off-diagonal contributions from both $\bm{r}$ and $n$:
\bea
{\cal _AH}(n,\bm{r},n',\bm{r}') &=&
\sum_{n''=1}^\infty \int d\bm{r}'' \; {\cal H}(n,\bm{r},n'',\bm{r}'')  
{\cal N}(n'',\bm{r}'',n',\bm{r}') \\  
&=& \left( E_{A-1}^{n} - \frac{\nabla^2_{r}}{2m} \right) 
{\cal N}(n,\bm{r},n',\bm{r}') 
+ \delta(\bm{r}-\bm{r}')\langle\Psi^n_{A-1}|\sum_{i=1}^{A-1} 
V(|\bm{r}_i - \bm{r}|)
| \Psi^{n'}_{A-1} \rangle  \nn\\ 
&& -(A-1) \int \prod_{i=2}^{A-1} d\bm{r}_i \; 
\Psi_{A-1}^{n*}(\bm{r}',\bm{r}_2, \ldots,\bm{r}_{A-1})  \nn\\ 
&& \times \left[V(|\bm{r'} - \bm{r}|) + \sum_{i=2}^{A-1} V(|\bm{r}_i - 
\bm{r}|)\right]
\Psi_{A-1}^{n'}(\bm{r},\bm{r}_2,\ldots,\bm{r}_{A-1}) \; .
\label{eq:hama}
\eea
Imposing antisymmetry via the projection operator associated
with ${\cal N}(n,\bm{r},n',\bm{r}')$ leads to exchange terms in the effective
potential.

For the special case of a one-body Hamiltonian, $H=\sum_{i=1}^A H(\bm{r}_i)$,
we obtain the Hamiltonian kernels ${\cal H}(n,\bm{r},n',\bm{r}') =
\delta_{nn'}\delta(\bm{r}-\bm{r}') \left(E_{A-1}^n + H(\bm{r})\right)$ and
${\cal _A H}(n,\bm{r},n',\bm{r}') = {\cal N} (n,\bm{r},n',\bm{r}')
\left(E_{A-1}^n + H(\bm{r})\right)$, and eq.~(\ref{eq:schro1b}) reduces to 
the single-particle Schr\"{o}dinger equation.  The kernel ${\cal N}
(n,\bm{r},n',\bm{r}')$ guarantees that only expansion coefficients originating
from an antisymmetric $A$-body wave function are considered, i.e. ${\cal N}
(n,\bm{r},n',\bm{r}')$ ensures that the associated single-particle orbitals are
not among the occupied states of the ($A$-1)-body system.  This can be
trivially taken into account in the calculations.  Nevertheless, even in a case
as simple as this one, ${\cal N} (n,\bm{r},n',\bm{r}')$ is not diagonal in
$n,n'$.

In order to facilitate working with the above (exact) set of coupled integral
equations, eq.~(\ref{eq:schro1b}), various approximations may be considered.
For example, the Hartree-Fock equations are recovered by taking the
$(A-1)$-body basis states, $\Psi^n_{A-1}$, to be Slater determinants
constructed from the expansion coefficients $\phi_n(\bm{r})$.  In contrast to
the approach presented in the previous section, the current method does not
require channel coupling to obtain the Fock terms.  Instead, the nonlocal
(exchange) terms are now explicitly present in the effective potential, as can
be seen in the last line of eq.~(\ref{eq:hama}).  Thus, the advantage of using
antisymmetric basis states --- coupled channels are not needed to include the
Fock term contributions --- is offset by additional complications in the
resulting equations of motion: the Hamiltonian kernel is no longer local and the
antisymmetry operator appears explicitly in eq.~(\ref{eq:schro1b}).  Note also
that in the Hartree-Fock approximation the channels are implicitly coupled
through the use of expansion coefficients in the $(A-1)$-body Slater
determinants.

Another approximation method leads to equations which were previously obtained
by Varga and Lovas {\em et al.}~\cite{Lovas} in the framework of the cluster 
model.  In
this approach, we ignore those terms in ${\cal _AH}(n,\bm{r},n',\bm{r}')$ and
${\cal N}(n,\bm{r},n',\bm{r}')$ that couple different values of the discrete
variable, i.e. contributions for which $n \neq n'$ holds.  The equations of
motion for the coefficients $\phi_n(\bm{r})$ then take the following form:
\bea
\int d\bm{r}'{\cal _AH}(n,\bm{r},n,\bm{r}') \phi_n(\bm{r}') 
= E \int d\bm{r}' {\cal N}(n,\bm{r},n,\bm{r}') \phi_n(\bm{r}') \; ,
\label{eq:lovasa}                                   
\eea
that is, we have effectively integrated out the coordinates of $(A-1)$
particles to obtain a set of one-body equations.  This equation is more general
than Hartree-Fock, but the Hartee-Fock approximation can be recovered by taking
the $(A-1)$-body wave functions to be Slater determinants composed from the
spectroscopic amplitudes.  We can then derive the one-body potential model,
eq.~(\ref{eq:1bSchro}), as in the previous section.  In these cases
antisymmetry can be exactly imposed through the use of Slater determinants and
the explicit projection operator, $\hat{{\cal N}}$, is not needed.

A second one-body equation can be obtained by taking the $(A-1)$-body wave
functions from an external source independent of the spectroscopic amplitudes.
This can go beyond Slater determinants and Hartree-Fock since the $(A-1)$-body
wave functions can include many-body correlations, in principle they can even
be exact. The quality of this approximation will depend on how well the
$(A-1)$-body wave functions are chosen.  In this approach the orbitals can be
considered one at a time. The price paid for this convenience is twofold.
First, the expressions for the Hamiltonian become more complicated, compare
eq.~(\ref{eq:ham}) and eq.~(\ref{eq:hama}).  Secondly, when we move beyond
Slater determinants, $\hat{{\cal N}}$ is explicitly required
and, moreover, its diagonal element, $\hat{{\cal N}}_{nn}$, is no
longer a projection operator (see Appendix B), so antisymmetry is not
explicit.

The one-body models associated with eqs.~(\ref{eq:1bSchro}) and
(\ref{eq:lovasa}) differ, but in most situations both lead to a decoupling of
the expansion coefficients.  The exceptions are the Hartree-Fock approximation
and other self-consistent models in which the $(A-1)$-body wave functions are
constructed from the expansion coefficients.  While the decoupling is in many
ways advantageous, it also implies that we have lost the information on the
relative normalizations of the $\phi_n(r)$.  If we assume that $A$ of the
coefficients are nonzero we may take them to be individually normalized to one.
This is consistent with the overall normalization condition $A =
\sum_{n'=1}^\infty \sum_{n=1}^\infty \int d\bm{r} d\bm{r}' \phi_n^*(\bm{r}) 
{\cal N} (n,\bm{r},n',\bm{r}') \phi_{n'}(\bm{r}')$.

In the work of Lovas {\em et al.}, eq.~(\ref{eq:lovasa}) was considered for the
case $n=n'=1$.  The Hamiltonian operator 
$\hat{h}=(\hat{{\cal N}}_{11})^{-1/2} \; {\cal _A}\hat{{\cal H}}_{11} \;
(\hat{{\cal N}}_{11})^{-1/2}$ and the wave function 
$\chi_l=(\hat{{\cal N}}_{11})^{1/2} \phi_1$ were defined.
These redefinitions have the advantage of leading to an equation which takes the 
same form as the standard Schr\"{o}dinger equation, 
$\hat{h} \chi_l=E\chi_l$. The $\chi_l$ provides a better approximation to the true 
spectroscopic amplitude than the model $\phi_1$ (the solution of 
eq.~(\ref{eq:lovasa})) since it includes the effect of $(\hat{{\cal N}}_{11})^{1/2}$, 
an approximate projection operator onto antisymmetric states. As noted previously, 
a projection operator may be needed and this is amplified in the discussion of 
ref.~\cite{Csoto00} in the next section. 
It is also $\chi_l$ (normalized to one) that in this approach should
be identified with the single-particle wave function, both because of this
better correspondence with the true spectroscopic amplitude and because of the
form of the equation which it satisfies.

\section{Spectroscopic Amplitudes and Reaction Rates}
\label{sec:Csoto}

In this section we discuss how the spectroscopic amplitudes can be used to
calculate reaction rates and elaborate on their physics content.  We show
that the spatial dependence of the spectroscopic amplitude and its norm, the
spectroscopic factor, describe different physical aspects of the many-nucleon
system, namely single-particle properties of the $A$-th particle and
distortions of the $(A-1)$-body system, respectively.  We make the connection
to the potential-model approach and find that both aspects are included, at
least approximately, when spectroscopic factors are employed to scale the wave
functions.  Since our findings contradict the conclusions of
ref.~\cite{Csoto00}, we explore the claims made in that paper.

To calculate the relevant reaction cross sections, we expand both the $A$-body
bound state, $|\psi_A(\bm{r}_1,\cdots,\bm{r}_A) \rangle$, and the wave
function in the incident channel, $|\psi^K_A(\bm{r}_1,\cdots,\bm{r}_A)
\rangle$, as in eq.~(\ref{eq:expn}). Here $K$ specifies the asymptotic momentum
of the incident particle relative to the ($A$-1)-body target nucleus.  The
corresponding expansion coefficients are given by $\phi_n(\bm{r}) = \sqrt{A}
\langle\Psi^{n,r}_A |\psi_A(\bm{r}_1,\cdots,\bm{r}_A) \rangle$ and
$\phi^K_n(\bm{r}) = \sqrt{A}
\langle\Psi^{n,r}_A |\psi^K_A(\bm{r}_1,\cdots,\bm{r}_A)
\rangle$, respectively. 
The expansion coefficient $\phi^K_1(\bm{r})$, associated with the continuum
wave function, is the optical model wave function~\cite{Feshbach,OptModel}.
Thus the formalism based on expansion coefficients is sufficiently general to
include both spectroscopic amplitudes and optical model wave functions.  It
will be useful whenever we are dealing with one-body operators.  The matrix
element for the one-body transition operator ${\cal O}(\bm{r})$ can then be
written as:
\bea
{\cal M} &\equiv& \langle \psi_A |\sum_{i=1}^A {\cal O}(\bm{r}_i)| \psi^K_A \rangle  
= A \langle \psi_A | {\cal O}(\bm{r}_A)|\psi^K_A  \rangle \\
&=& \sum_{n=1}^\infty \int d\bm{r} \phi^*_n(\bm{r}){\cal O}(\bm{r})
\phi^K_n(\bm{r})    \; .
\label{eq:matrix}                  
\eea
To make the connection with the potential model we separate the spatial
dependence of the spectroscopic amplitude and its normalization, $\sqrt{S_n}$,
as follows:
\bea
\phi_n(\bm{r})=\sqrt{S_n} \tilde\phi_n(\bm{r}) \; ,
\label{eq:SepSpatNorm} 
\eea
where $\int d\bm{r} | \tilde{\phi}_n(\bm{r}) |^2 = 1$.  In
Section~\ref{sec:sa}, we have shown that the potential-model wave functions
approximate the spectroscopic amplitudes.  Since the norm of
$\tilde{\phi}_n(\bm{r})$ is one, whereas that of $\phi_n(\bm{r})$ is $S_n$, the
$\tilde{\phi}_n(\bm{r})$, rather than the $\phi_n(\bm{r})$ should be identified
with the potential-model wave functions.  There is no equivalent normalization
factor associated with the scattering state; $\psi^K_A$ is normalized
asymptotically.

The transition matrix element given in eq.~(\ref{eq:matrix}) can now be written
as ${\cal M} = \sum_{n=1}^\infty \sqrt{S_n} \int d\bm{r} \tilde{\phi}^*_n(\bm{r}) 
{\cal O}(\bm{r}) \phi^K_n(\bm{r})$.  Both the $^7$Be$(p,\gamma)^8$B and
$^{16}$O$(p,\gamma)^{17}$F$^*$ reactions at threshold are peripheral, i.e. the
capture processes take place at large distances from the center of the target
nucleus, which is in its ground state.  In such situations, as for all direct
capture reactions, only the first expansion coefficient for $\psi^K_A$
contributes and the matrix element reduces to ${\cal M} \rightarrow \sqrt{S_1}
\int d\bm{r} \tilde{\phi}^*_1(\bm{r}) {\cal O}(\bm{r}) \phi^K_1(\bm{r})$. 
Since the cross section is proportional to $|{\cal M}|^2$, the spectroscopic
factor associated with the ground state of the ($A$-1)-body system, $S_1$,
occurs linearly in the expression for the reaction rate.  Thus we conclude that
when a potential-model wave function (or any other function normalized to one)
is used to calculate the transition matrix element ${\cal M}$, the resulting
cross section needs to be multiplied by the associated spectroscopic factor in
order to account for (some of) the many-body correlations in the nuclei
involved. The implications of this for the asymptotic normalization are
discussed at the end of this section.

The separation introduced in eq.~(\ref{eq:SepSpatNorm}) is motivated by the
realization that the spatial dependence of the spectroscopic amplitude
$\phi_n(\bm{r}_A)$, and its norm, the spectroscopic factor $S_n$, describe
different physical properties of the nuclear many-body system.  The former is
related to the shape of the single-particle orbital of the $A$-th nucleon in
the system, and can therefore be expressed through the normalized amplitude
$\tilde{\phi}_n(\bm{r})$.  The latter provides a measure of the structural
similarity of the $n$-th excited $(A-1)$-body state and an $(A-1)$-body
subcluster of the larger system.  Equivalently, the set of spectroscopic
factors associated with the expansion of $\psi_A$, eq.~(\ref{eq:expn}), can be
viewed as describing the distortion of the $(A-1)$-body core due to the
presence of an extra nucleon.  This can be seen, for example, by casting $S_n$
into the following form:
\bea
S_n & = & \int dr \left|\phi_n(r)\right|^2  \\
    & = & A \int \left(\prod_{i=1}^{A-1} dr_i \right)
           \int \left(\prod_{i=1}^{A-1} dr'_i \right)
           \Psi_{A-1}^{n*}(r_1,\cdots,r_{A-1})
    \Psi_{A-1}^n(r'_1,\cdots,r'_{A-1})\nonumber  \\ 
&& \times \left[ \int dr \psi^*_A(r_1,\cdots,r_{A-1},r)
    \psi_A(r'_1,\cdots,r'_{A-1},r) \right]         \; . 
\eea
In the last line (expression in square brackets) we have integrated out the
dependence on the $A$-th particle.  We are left with expressions involving wave
functions of the $(A-1)$-body system.  The extra particle's influence is still
present in the modification it has induced in the $(A-1)$-body cluster.  Upon
decoupling the equations of motion for the expansion coefficients, as required
in our derivation of the one-body models, this information on the last
particle's influence is lost, the many-body correlations contained in the
integrals $S_n$ disappear and the spectroscopic factors become one.

The spectroscopic factors carry information both on the dynamical distortions
induced by the interaction between the ($A$-1)-body system and the extra
particle and on antisymmetry effects.  Specifically, as shown in
Appendix~\ref{sec:App_norm}, the spectroscopic factor can be written as:
\begin{eqnarray}
S_n=N_n \langle\psi^{\parallel,n}_A |\psi^{\parallel,n}_A\rangle \; ,
\label{eq:nn}
\end{eqnarray}
where $\langle\psi^{\parallel,n}_A |\psi^{\parallel,n}_A\rangle$ reflects the
influence of the distortions, and the normalization factor $N_n$ contains
antisymmetry effects.  Specifically, $N_n$ keeps track of the requirement that
the nucleon orbitals in the $n$-th excited $(A-1)$-body state must be
orthogonal to the orbital $\tilde\phi_n(\bm{r})$ of the $A$th particle (for
details see eq.~(\ref{eq:tnorm})).  When center-of-mass corrections are taken
into account, $N_n$ may be greater than one, otherwise it is less than or equal
to one.  The matrix element $\langle\psi^{\parallel,n}_A |\psi^{\parallel,n}_A
\rangle$ is always less than or equal to one.  Consequently, when
center-of-mass corrections are ignored, we have the restriction $S_n \leq 1$
for the spectroscopic factor.  The influence of these corrections are discussed
in detail in Appendix~\ref{App:COMcorr}.

At this point we would like to reassert that it is correct and necessary to
include spectroscopic factors in potential-model cross sections of nuclear
capture reactions.  This procedure was questioned in ref.~\cite{Csoto00}.  The
author of that paper argues that the short-range correlations contained in the
spectroscopic factor should modify the bound-state wave function only in the
nuclear interior and have no effect on the asymptotic behavior.  Since
multiplying the potential model wave function by $\sqrt{S_n}$, however, affects
its overall normalization, including in the tail region, the usual procedure of
treating microscopic correlations in the potential model, viz. through
spectroscopic factors, is pronounced to be incorrect.  To illustrate his point,
the author compares a wave function $\chi_c(\bm{r})$, which describes the
relative motion of $^7$Be and $p$ in $^8$B, to the spectroscopic amplitude
function of the $^7$Be+$p$ configuration in $^8$B [Note that this $\chi_c$ is
different from the wave function $\chi_l$ introduced by Lovas {\em et
al.}~\cite{Lovas} and discussed in the previous section].  The wave functions
$\chi_c$ considered individually do not contain Pauli effects but, when used in
cluster-model calculations, appear behind an antisymmetrization operator.  The
spectroscopic amplitudes, on the other hand, are calculated from properly
antisymmetrized wave functions.  The two functions are shown to agree with each
other and with the appropriately normalized Coulomb-Whittaker function in the
asymptotic region, but they differ at small radii, as can be seen in Fig.~2 of
ref.~\cite{Csoto00}.  Their difference is interpreted as a measure of the Pauli
effects.  Since multiplying $\chi_c(\bm{r})$ by a spectroscopic factor would
affect both its short-range and asymptotic behavior, the author concludes that
this cannot be the proper procedure for incorporating microscopic substructure
effects in potential model calculations.

The argument presented in ref.~\cite{Csoto00} is not correct.  The function
$\chi_c(\bm{r})$ does not, in general, correspond to a potential-model wave
function. First, unlike the one-body wave functions we have considered, it
already includes the spectroscopic factor through the normalization of the
$A$-body wave function.  Moreover, very little physics can be associated with
$\chi_c$ outside the context of its usual use as a component in a properly
antisymmetrized cluster-model wave function.  To show this, we consider an
arbitrary product state $\psi_A^P(\bm{r}_1,\cdots,\bm{r}_A) = \varphi(\bm{r}_A)
\psi_{A-1}^P(\bm{r}_1, \cdots,\bm{r}_{A-1})$.
In the nonantisymmmetrized basis, it has expansion coefficients:  
\bea
\phi^P_n(\bm{r}) = \sqrt{A} \langle\Psi^{n,r}_A|\psi^P_A\rangle 
= \varphi(\bm{r}) \sqrt{A} \langle\Psi^n_{A-1} |\psi_{A-1}^P\rangle \; ,
\label{eq:prod}
\eea 
and the associated antisymmetrized, i.e. physical state, has expansion
coefficients $\phi_n(\bm{r}) = (1/A) \sum_{n'=1}^\infty\int d\bm{r} {\cal
N}(n,\bm{r},n',\bm{r}') \phi^P_{n'} (\bm{r})$.  In the product given above,
$\varphi(\bm{r})$ can be taken to correspond to the relative-motion function
$\chi_c(\bm{r})$ of ref.~\cite{Csoto00}.  As discussed in the previous section,
$\hat{{\cal N}}$ enforces antisymmetry by projecting onto a
completely antisymmetric state.  Thus, the choice of $\varphi(\bm{r})$ is
somewhat arbitrary, since many functions lead to the same physical state.  We
may, for example, consider the case in which $\psi_{A-1}^P$ and
$\Psi^{n}_{A-1}$ are Slater determinants and $\psi_{A-1}^P$ = $\Psi^{1}_{A-1}$.
If we then take $\varphi(\bm{r})$ to be orthogonal to the orbitals in
$\psi_{A-1}^P$, the antisymmetry requirement will turn the product into a
Slater determinant and we obtain $\phi_1(\bm{r})=\phi^P_1(\bm{r})/\sqrt{A} =
\varphi(\bm{r})$.  On the other hand, if $\varphi(\bm{r})$ is not orthogonal to
the occupied orbitals, the non-orthogonal components will be projected out as
well.  In the extreme case of $\varphi({\bm r})$ being a linear combination of
the occupied states, $\phi_1(\bm{r})$ is zero.  From these considerations we
conclude that the difference between $\varphi(\bm{r})$ and the associated
physical state has no particular significance.  The effect that
antisymmetrization has on $\varphi({\bm r})$ is not pertinent to the one-body
models considered in this paper since both already include antisymmetry, at
least approximately. (The effect of antisymmetry on the shape of the
spectroscopic amplitudes is distinct from the effect of antisymmetry contained
in the $N_n$ of eq.~(\ref{eq:nn}).)  A more useful comparison would be between
the spectroscopic amplitude and $(\hat{{\cal N}}_{11})^{1/2} \varphi(\bm{r}')$, 
since this would measure the influence of the off-diagonal matrix elements of 
$\hat{{\cal N}}$ and thus test the validity of the single-particle
model based on eq.~\ref{eq:lovasa}). Neither comparison clarifies the role of
the spectroscopic factors since, as previously noted, the spectroscopic factor
is contained in both $\chi_c$ and the spectroscopic amplitudes.

At this point it is useful to return to the asymptotic normalization, $A_{nc}$,
of the spectroscopic amplitude.  For the low-energy $^7$Be$(p,\gamma)^8$B and
$^{16}$O$(p,\gamma)^{17}$F$^*$ reactions, for example, the capture occurs at
large radii.  Thus the asymptotic normalization of the spectroscopic amplitude
is sufficient for describing the bound state in the physically relevant region.
This is in line with the conclusions of refs.~\cite{Muk01,Muk99}.  As explained
above, there are two different physical effects included in the spectroscopic
amplitude and hence in its asymptotic normalization: one is related to the
distortions of the $(A-1)$-body cluster due to the presence of an additional
nucleon and is contained in the spectroscopic factor and the other is related
to the single-particle properties of that extra particle and is described by
the spatial dependence of the amplitude. Both are needed.

\section{Conclusions}
\label{sec:Concl}

Spectroscopic amplitudes play a central role in the description of
single-particle transfer reactions such as radiative nucleon capture.  The
amplitudes contain both single-particle and many-nucleon aspects of the nuclear
many-body problem and can be, in principle, obtained from a fully microscopic
model.  We have presented two alternative approaches, each based on a set of
coupled-channels equations, and have shown how one-body approximations can be
derived in a systematic manner.  We obtained two different single-particle
models.  In both cases, the single-particle wave function was found to be an
approximation to the spectroscopic amplitude but normalized to one rather than
to the spectroscopic factor.  The quality of the one-body approximation depends
on how well it describes the spatial dependence of the spectroscopic amplitude.
However, even the best single-particle wave function will miss the
spectroscopic factor.

The first one-body approximation considered here results in $A$ noninteracting
particles in a single-particle potential.  The simplification occurs since the
equations for the spectroscopic amplitudes are no longer coupled; instead, we
have $A$ independent equations.  Since we still have $A$ particles and $A$
orbitals, we can construct an antisymmetric $A$-body wave function by taking it
to be a Slater determinant.  In this model, it is redundant to explicitly
enforce antisymmetry, e.g.  through projection operators.  The quality of this
approach depends on how well the single-particle potential is chosen and how
well the single-particle wave functions reproduce the shape of the
spectroscopic amplitudes.

The second one-body model presented here is usually derived within the
generator-coordinate formalism.  We obtained it by truncating a set of coupled
integral equations.  In this approach, antisymmetry is (approximately) imposed
for each orbital separately through (approximate) projection operators. Unlike
the first one-body approximation it can include many-particle correlations in
the $(A-1)$-body subsystem.  In principle, this approach is also simpler since
we do not need to consider all the orbitals together.  The price we pay is that
the potential cannot be approximated as simply since it implicitly contains a
projection into antisymmetric states.  There is also an explicit projection
operator that must be approximated.  The quality of this approach depends on
the choice of the $(A-1)$-body wave functions and the importance of the channel
coupling.

It is important to realize that there is not one unique single-particle model.
Different one-body approximations to the nuclear many-body problem exist and
can be derived independently of each other.  The resulting single-particle
models may differ in subtle but crucial details and should therefore not be
confused with each other.  We have derived two such models and discussed their
relation to each other and distinctions between them.  In particular, the very
different techniques for including antisymmetry should be noted.

The spectroscopic factor, in its simplest form, reflects the partition
probability of the $A$-body system into smaller clusters with allowance for
antisymmetry effects. In the present paper it is cast in a complimentary
light. It presents itself as a manifestation of the distortion of the
$(A-1)$-body system due to the presence of the $A$th particle. This distortion
is both dynamical due to the interactions and kinematical due to
antisymmetry. Since these are pure many-body effects they are, by definition,
absent from one-body approximations. When center-of-mass corrections are
included, the spectroscopic factors can be greater than one without violating
the Pauli principle. Otherwise they must be less than or equal to one.

A full calculation must include both the one-body and many-body effects.  As
the present paper emphasizes, this can be accomplished through the use of
spectroscopic amplitudes.  Obtaining these amplitudes, however, requires the
full solution of the coupled equations presented here or a fully microscopic
model.  While this is still a distant goal, some recent fully microscopic
approaches show promising results for low-mass systems~\cite{Kamada}.
Furthermore, models such as the shell model, the continuum shell model, or the
cluster model include some many-nucleon correlations and provide reasonable
approximations to the full problem.  Since many-body effects are not contained
in one-body models, cross sections calculated in this framework need to include
the spectroscopic factor.  For processes that are strongly peaked in the tail
region, like the low-energy $^7$Be$ (p,\gamma)^8$B and $^{16}$O$
(p,\gamma)^{17}$F$^*$ reactions, the one-body and many-body effects can be
combined into a single parameter --- the asymptotic normalization coefficient.

\acknowledgements{One of the authors (H.S.S.) is grateful to the Theory 
Group at TRIUMF for their hospitality.  The work presented here was supported
by funding from the Natural Sciences and Engineering Research Council of
Canada. TRIUMF receives federal Canadian funding via a contribution agreement
through the National Research Council of Canada.}

\appendix
\section{The Bases}
\label{sec:App_bases}

We begin by defining a basis $\{ \Psi^n_{A-1} \}_{n=1,2,\ldots}$ for ${\rm
H}_{A-1}^{\cal A}$, the space of completely antisymmetric $(A-1)$-body wave
functions $\psi_{A-1}(\bm{r}_1,\cdots,\bm{r}_{A-1})$.  The basis states are
orthonormal,
\bea
\int \prod_{i=1}^{A-1}d\bm{r}_i
     \Psi_{A-1}^{n*}(\bm{r}_1,\cdots,\bm{r}_{A-1})
     \Psi_{A-1}^{n'}(\bm{r}_1,\cdots,\bm{r}_{A-1}) =
     \delta_{nn'}  \; , 
\eea
and complete in ${\rm H}_{A-1}^{\cal A}$, that is:
\bea
 \int \prod_{i=1}^{A-1}d\bm{r}'_i
\sum_{n=1}^{\infty}
  \Psi_{A-1}^n(\bm{r}_1,\cdots,\bm{r}_{A-1})
  \Psi_{A-1}^{n*}(\bm{r}'_1,\cdots,\bm{r}'_{A-1}) \;
  \psi_{A-1}(\bm{r}'_1,\cdots,\bm{r}'_{A-1})
&&\nonumber\\ 
  = \psi_{A-1}(\bm{r}_1,\cdots,\bm{r}_{A-1})      
\eea
holds for any $\psi_{A-1} \in {\rm H}_{A-1}^{\cal A}$.  Specifically, for the
sake of convenience, we choose $\Psi^n_{A-1}$ which are eigenstates of the
$(A-1)$-body Hamiltonian $H_{A-1}=-\sum_{i=1}^{A-1}\frac{\nabla^2_{r_i}}{2m_i}
+ \frac{1}{2} \sum_{i,j=1}^{A-1} V(|\bm{r}_i - \bm{r}_j|)$.  To do so, we have
to include both bound and scattering states.  The superscript $n$ labels the
discrete as well as the continuous spectrum of $H_{A-1}$.

We now consider two different $A$-body spaces. The first, denoted by ${\rm
H}_{A}$, is spanned by
\bea
\Psi^{n,r}_A(\bm{r}_1,\cdots,\bm{r}_A) \equiv
\Psi^n_{A-1}(\bm{r}_1,\cdots,\bm{r}_{A-1})
\delta(\bm{r}-\bm{r}_A) \; ,
\label{eq:bd} 
\eea
where $\bm{r}$ is a continuous parameter. The $\Psi^{n,r}_A$ are orthonormal
with respect to both $n$ and $\bm{r}$:
\bea
\int \prod_{i=1}^A d\bm{r}_i        
     \Psi_A^{n,r*}(\bm{r}_1,\cdots,\bm{r}_A)
     \Psi_A^{n',r'}(\bm{r}_1,\cdots,\bm{r}_A) =
     \delta_{nn'}\delta(\bm{r}-\bm{r}') \; , 
\eea
and the completeness condition for this basis is given by:
\bea
\int \prod_{i=1}^A d\bm{r}'_i
\left(\sum_{n=1}^{\infty} \int d\bm{r} \;
  \Psi_A^{n,r}(\bm{r}_1,\cdots,\bm{r}_A)
  \Psi_A^{n,r*}(\bm{r}'_1,\cdots,\bm{r}'_A) \right)
  \psi_A(\bm{r}'_1,\cdots,\bm{r}'_A) && \nonumber\\ 
  = \psi_A(\bm{r}_1,\cdots,\bm{r}_A)  \; ,     
\eea
where $\psi_A \in {\rm H}_A$.  The space ${\rm H}_A$ is a direct sum of the
subspaces ${\rm H}_A^{\cal A}$ and ${\rm H}_A^{\cal M}$ which contain,
respectively, totally antisymmetric and mixed-symmetry $A$-body states.  The
latter are antisymmetric in the first $A$-1 coordinates and symmetric with
respect to exchanges between the $A$-th nucleon and any other particle.  For
$A$=2, the space ${\rm H}_{A=2}^{\cal M}$ is completely symmetric.

An arbitrary wave function $\psi_A \in {\rm H}_A$ can thus be written as the
sum of an antisymmetric and a mixed-symmetric component:
\bea
\psi_A(\bm{r}_1,\cdots,\bm{r}_A) &=&
  \frac{\cal A}{\sqrt{A}}
\psi_A(\bm{r}_1,\cdots,\bm{r}_A) 
  + \left(1-\frac{\cal A}{\sqrt{A}}\right)
\psi_A(\bm{r}_1,\cdots,\bm{r}_A) \\
  &=& \psi_A^{\cal A}(\bm{r}_1,\cdots,\bm{r}_A) 
  + \psi_A^{\cal M}(\bm{r}_1,\cdots,\bm{r}_A) \; ,
\eea
where $\psi_A^{\cal A} \in {\rm H}_A^{\cal A}$ and $\psi_A^{\cal M} \in {\rm
H}_A^{\cal M}$, and $\cal A$ denotes an `intercluster' antisymmetrization
operator, which antisymmetrizes between the $A$-th coordinate and the remaining
$A$-1 coordinates.  $\cal A$ is normalized by the condition ${\cal
A}^2=\sqrt{A}\cal A$.  Since $({\cal A}$ $/{\sqrt{A}}) \left( 1-{\cal A}/{
\sqrt{A}}\right)=0$, the two subspaces are orthogonal to each other, that is:
\bea
\int \prod_{i=1}^A d\bm{r}_i \;
\psi_A^{\cal A*}(\bm{r}_1,\cdots,\bm{r}_A) \;
\psi_A^{\cal M}(\bm{r}_1,\cdots,\bm{r}_A)&=&0 \; .
\eea
Furthermore, a symmetric operator $\hat{{\cal O}}^{\cal S}$ cannot connect the
two subspaces:
\bea
\int \prod_{i=1}^A d\bm{r}_i \;
\psi_A^{{\cal A}*}(\bm{r}_1,\cdots,\bm{r}_A) \;
\hat{{\cal O}}^{\cal S} \;
\psi_A^{\cal M}(\bm{r}_1,\cdots,\bm{r}_A)&=&0 
\eea
This includes the case where $\hat{{\cal O}}^{\cal S}= H_A$, where
$H_A$ is the $A$-body Hamiltonian.

The space ${\rm H}_A^{\cal A}$ is spanned by
\bea
  _{\cal A}\Psi_A^{n,r}(\bm{r}_1,\cdots,\bm{r}_A)
\equiv {\cal A} \Psi_A^{n,r}
  = {\cal
A}[\Psi_{A-1}^n(\bm{r}_1,\cdots,\bm{r}_{A-1})\delta(\bm{r}-\bm{r}_A)]
\eea
or, equivalently,
\bea
  |_{\cal A}\Psi_A^{n,r}\rangle = a^\dagger(\bm{r})
\; |\Psi_{A-1}^n\rangle \;
   ,
\label{eq:sqbasis}
\eea
where $a^\dagger(\bm{r})$ creates a nucleon at position $\bm{r}$ and we have
used the convention $\psi_A(\bm{r}_1,\cdots,\bm{r}_A) = \langle
\bm{r}_1,\cdots,\bm{r}_A | \psi_A \rangle$. The creation and annihilation
operators $a^\dagger(\bm{r})$ and $a(\bm{r})$ obey the usual anticommutation
relations, which ensure that the right-hand side of eq.~(\ref{eq:sqbasis}) is
totally antisymmetric.  The completeness condition for the basis $\{ _{\cal
A}\Psi_A^{n,r} \}$ takes the form:
\bea
\int \prod_{i=1}^A d\bm{r}'_i \left(\frac{1}{A}\sum_{n=1}^{\infty} 
  \int d\bm{r} \; _{\cal A}\Psi_A^{n,r}(\bm{r}_1,\cdots,\bm{r}_A) \;
  _{\cal A}\Psi_A^{n,r*}(\bm{r}'_1,\cdots,\bm{r}'_A)\right) 
  \psi_A(\bm{r}'_1,\cdots,\bm{r}'_A) && \label{eq:normbaa} \\ 
   = \psi_A(\bm{r}_1,\cdots,\bm{r}_A) \; ,  \nn 
\eea
where $\psi_A$ is a fully antisymmetric $A$-body wave function from ${\rm
H}_A^{\cal A}$.  When $\psi_A$ in eq.~(\ref{eq:normbaa}) is replaced by a state
from ${\rm H}_A^{\cal M}$, the right-hand side of the equation vanishes.  
Thus ${\cal P}_{\cal A} \equiv \frac{1}{A} \sum_{n=1}^{\infty} \int d\bm{r} \; 
|_{\cal A}\Psi_A^{n,r} \rangle \langle _{\cal A}\Psi_A^{n,r}|$ is a projection 
operator which projects states $\psi_A \in {\rm H}_A$ onto their antisymmetric 
component $\psi_A^{\cal A} \in {\rm H}_A^{\cal A}$.
That ${\cal P}_{\cal A}^2 = {\cal P}_{\cal A}$ holds can be shown by using
eq.~(\ref{eq:overcompbas}) below.	

The advantages of using totally antisymmetric basis states are obvious.  The
disadvantages of employing this basis lies in the fact that the states are no
longer orthonormal.  Instead, we have:
\bea
\int \prod_{i=1}^A d\bm{r}_i{} \;
     _{\cal A}\Psi_A^{n,r*}(\bm{r}_1,\cdots,\bm{r}_A) \;
     _{\cal A}\Psi_A^{n',r'}(\bm{r}_1,\cdots,\bm{r}_A) =
     {\cal N}(n,\bm{r},n',\bm{r}')   \; .     
\eea
The norm operator, $\hat{{\cal N}}$, and its kernel, ${\cal N}(n,\bm{r},n',\bm{r}')$, 
have various interesting properties and are discussed in Appendix~\ref{sec:App_normOp}.

The basis states $_{\cal A}\Psi_A^{n,r}$ are not linearly independent, but are
related to each other through the norm operator, $\hat{{\cal N}}$, as follows:
\bea 
_{\cal A}\Psi_A^{n,r}(\bm{r}_1, \cdots,\bm{r}_{A}) &=& 
\frac{1}{A} \sum_{n'=1}^\infty \int d\bm{r}' {\cal N} (n,\bm{r},n', \bm{r}') 
_{\cal A}\Psi_A^{n',r'} (\bm{r}_1, \cdots,\bm{r}_{A})  \; .
\label{eq:overcompbas}
\eea 
In fact, the $\{ _{\cal A}\Psi_A^{n,r} \}$ basis is overcomplete and, at least 
in the case where the $\Psi^n_{A-1}$ are Slater determinants, spans the space 
${\rm H}^{\cal A}_A$ $A$ times.  This accounts for the factor $\frac{1}{A}$ in the 
first line of eq.~(\ref{eq:expn}), and in the completeness relation,
eq.~(\ref{eq:normbaa}).

An arbitrary antisymmetric $A$-body wave function $\psi_A(\bm{r}_1,
\cdots,\bm{r}_A)$ can now be expanded in one of the above bases.  Using the set
$\{ \Psi_A^{n,r} \}$, we have:
\bea
\psi_A(\bm{r}_1,\cdots,\bm{r}_A) &=& 
   \frac{1}{\sqrt{A}} \sum_{n=1}^{\infty} \int
d\bm{r} \; 
   \Psi_A^{n,r}(\bm{r}_1,\cdots,\bm{r}_A)
\phi_n(\bm{r}) \; , 
\label{eq:b1a}
\eea 
with expansion coefficients
\bea
\phi_n(\bm{r}) &=& \sqrt{A}\int \prod_{i=1}^A d\bm{r}_i \;
  \Psi_A^{n,r*}(\bm{r}_1,\cdots,\bm{r}_A) \psi_A(\bm{r}_1,\cdots,\bm{r}_A) \; .
\label{eq:b1b}  
\eea
Alternatively, we can use the antisymmetric basis, $\{ _{\cal A}\Psi_A^{n,r} \}$, 
to write:
\bea
\psi_A(\bm{r}_1,\cdots,\bm{r}_A) &=&
\frac{1}{A}\sum_{n=1}^{\infty} 
   \int d\bm{r}\; _{\cal
A}\Psi_A^{n,r}(\bm{r}_1,\cdots,\bm{r}_A) 
   \phi_n(\bm{r}) \; ,
\label{eq:b2a}                       
\eea
where
\bea
\phi_n(\bm{r})=\int \prod_{i=1}^A d\bm{r}_i \; 
   _{\cal A}\Psi_A^{n,r*}(\bm{r}_1,\cdots,\bm{r}_A)
   \psi_A(\bm{r}_1,\cdots,\bm{r}_A) \; . 
\label{eq:b2b}   
\eea
Equation~(\ref{eq:b2a}) follows by applying the antisymmetrization operator to
eq.~(\ref{eq:b1a}) and eq.~(\ref{eq:b2b}) can be derived from
eq.~(\ref{eq:b1b}) by using the identity $\psi_A(\bm{r}_1,\cdots,\bm{r}_A) =
({\cal A}/\sqrt{A}) \; \psi_A(\bm{r}_1,\cdots,\bm{r}_A)$, which holds for totally
antisymmetric $A$-body states, and the Hermitean properties of $\cal A$.  Thus,
the coefficients are the same in both expansions.

 From eq.~(\ref{eq:b1a}), it follows that is also possible to write the wave
function $\psi_A$ as
\bea
\psi_A(\bm{r}_1,\cdots,\bm{r}_A) &=&
\frac{1}{\sqrt{A}} \sum_{n=1}^{\infty} \;
   \Psi_{A-1}^n(\bm{r}_1,\cdots,\bm{r}_{A-1}) \;
\phi_n(\bm{r}_A) \; .
\label{eq:b3a}
\eea 
The coefficients $\phi_n(\bm{r})$ are identical to those in the previous
expansions. From eqs.~(\ref{eq:sqbasis}) and (\ref{eq:b2b}), one infers that
they take the following form:
\bea
\phi_n(\bm{r}_A) &=& \langle \Psi_{A-1}^n | a(\bm{r})
| \psi_A  \rangle \; .
\label{eq:b3b}
\eea

When $\psi_A(\bm{r}_1,\cdots,\bm{r}_A)$ denotes a bound state and the
$\Psi_{A-1}^n$, which occur in the definitions of both $\Psi_A^{n,r}$ and
$_{\cal A}\Psi_A^{n,r}$, are eigenstates of the $(A-1)$-body system, then the
$\phi_n(\bm{r})$ are the {\em spectroscopic amplitudes} and the associated
integrals $S_n \equiv \int dr \left|\phi_n(\bm{r})\right|^2$ are the {\em
spectroscopic factors}. The $A$-dependent normalization factors in the above
equations are included so that $\int \prod_{i=1}^A d\bm{r}_i \left|
\psi_A(\bm{r}_1,\cdots,\bm{r}_A)\right|^2=1$ holds,
as well as
\bea
\sum_{n=1}^\infty \int d\bm{r}
\left|\phi_n(\bm{r})\right|^2 = A \; , 
\label{eq:pnorm}
\eea
in accordance with the conventional normalization of the spectroscopic factors.
This last equation follows by squaring eq.~(\ref{eq:b1a}), integrating over the
coordinates, and using the completeness of the $\Psi_A^{n,r}$.

\section{The norm operator}
\label{sec:App_normOp}

The norm operator for the antisymmetric basis $\{ _{\cal A}\Psi_A^{n,r} \}$, 
$\hat{{\cal N}}$, and its kernel, $ {\cal N}(n,\bm{r},n',\bm{r}')$, have many 
interesting properties.  To start with, the kernel can be written in several 
equivalent forms:
\bea
{\cal N}(n,\bm{r},n',\bm{r}') 
&=& 
 \int \prod_{i=1}^A d\bm{r}_i \; _{\cal A}\Psi_A^{n,r*}(\bm{r}_1,\cdots,\bm{r}_A) 
    _{\cal A}\Psi_A^{n',r'}(\bm{r}_1,\cdots,\bm{r}_A)
\label{eq:norm2} \\
&=&
  \sqrt{A} \int \prod_{i=1}^A d\bm{r}_i \;
  \Psi_A^{n,r*}(\bm{r}_1,\cdots,\bm{r}_A) 
     _{\cal A}\Psi_A^{n',r'}(\bm{r}_1,\cdots,\bm{r}_A) 
\label{eq:norm3} \\
&=&
  \sqrt{A} \int \prod_{i=1}^A d\bm{r}_i \;
  \Psi_A^{n,r*}(\bm{r}_1,\cdots,\bm{r}_A) 
     {\cal A}\Psi_A^{n',r'}(\bm{r}_1,\cdots,\bm{r}_A)
\label{eq:norm4} \\
&=&
   A \int \prod_{i=1}^A d\bm{r}_i \; \Psi_A^{n,r*}(\bm{r}_1,\cdots,\bm{r}_A)
     \frac{\cal A}{\sqrt{A}}\Psi_A^{n',r'}(\bm{r}_1,\cdots,\bm{r}_A)
\label{eq:norm5} \\
&=& 
   \langle \Psi_{A-1}^n| a(\bm{r}) a^\dagger(\bm{r}') |\Psi_{A-1}^{n'}\rangle
   \; . 
\label{eq:norm1}
\eea
We see that ${\cal N}(n,\bm{r},n',\bm{r}')$ is not only the kernel of the norm 
operator for $\{ _{\cal A}\Psi_A^{n,r} \}$, eq.~(\ref{eq:norm2}), but is also 
proportional to the overlap of an element from $\{ _{\cal A}\Psi_A^{n,r} \}$ with 
an element from the non-antisymmetrized basis $\{ \Psi_A^{n,r} \}$, 
eq.~(\ref{eq:norm3}).
Furthermore, ${\cal N}(n,\bm{r},n',\bm{r}') / \sqrt{A}$ is the matrix element
of the `intercluster' antisymmetrization operator, ${\cal A}$, in the basis $\{
\Psi_A^{n,r} \}$, eq.~(\ref{eq:norm4}), or -- equivalently -- ${\cal
N}(n,\bm{r},n',\bm{r}') / A$ is the matrix element of the projection operator
${\cal A}/\sqrt{A}$ in that same basis, eq.~(\ref{eq:norm5}).  Finally, we can
write ${\cal N}(n,\bm{r},n',\bm{r}')$ as the matrix element of $ a(\bm{r})
a^\dagger(\bm{r}')$ in the ($A$-1)-body basis, eq.~(\ref{eq:norm1}), where
$a^\dagger(\bm{r})$ and $a(\bm{r})$ create and annihilate, respectively, a
nucleon at position $\bm{r}$.

Since ${\cal N}(n,\bm{r},n',\bm{r}')$ is proportional to a projection operator,
it has no inverse.  However, its square root -- in the sense of a matrix
operation -- can be given.  It is simply the matrix element of the
antisymmetrization operator ${\cal A}$:
\bea
\sqrt{{\cal N}(n,\bm{r},n',\bm{r}')}
= \int \prod_{i=1}^A d\bm{r}_i \; \Psi_A^{n,r*}(\bm{r}_1,\cdots,\bm{r}_A)
     {\cal A}\Psi_A^{n',r'}(\bm{r}_1,\cdots,\bm{r}_A)  \; .
\eea 
To see the projection operator nature of $\hat{{\cal N}}$ more
directly, one can multiply eq.~(\ref{eq:b1a}) by $_{\cal A}\Psi_A^{n',r'}
(\bm{r}_1,\cdots,\bm{r}_A)$ and integrate over the coordinates.  If $\psi_A$ is
completely antisymmetric, we obtain the following equation for the expansion
coefficients:
\bea
\phi_n(\bm{r}) = \frac{1}{A} \sum_{n'=1}^\infty \int dr'\; 
   {\cal N}(n,\bm{r},n',\bm{r}') \phi_{n'}(\bm{r}')  \; .
\label{eq:Nprojphia} 
\eea
For a mixed-symmetric state $\psi_A$, on the other hand, we find:
\bea
0 = \frac{1}{A}\sum_{n'=1}^\infty \int d\bm{r}'\; 
{\cal N}(n,\bm{r},n',\bm{r}') \phi_{n'}(r')  \; ,
\label{eq:Nprojphib} 
\eea
i.e. $\hat{{\cal N}}/A$, when acting on a set of expansion
coefficients $\phi_n(\bm{r})$, behaves like a projection operator: It returns
the coefficients $\phi_n(\bm{r})$ of an antisymmetric state and yields zero
when the $\phi_n(\bm{r})$ correspond to a mixed-symmetric state.  It follows
immediately that in a restricted space of coefficients which originate from a
completely antisymmetric wave function, $\hat{{\cal N}}/A$
becomes the identity matrix.  In this restricted subspace, the functions
$_{\cal A}\Psi_A^{n',r'}/\sqrt{A}$ act in many ways as if they were
orthonormal.

If we expand an excited state of the $A$-body system,  $\Psi_A^k(\bm{r}_1,
\cdots,\bm{r}_A)$, as in eq.~(\ref{eq:expn}), we obtain expansion coefficients
$\phi^{A,k}_n(\bm{r}) \equiv \langle\Psi_{A-1}^{n}|a({\bm r}) | \Psi_{A}^k
\rangle$.  For $k=1$, $\Psi_A^{k=1}$ describes the ground state of the
$A$-nucleon system and the $\phi^{A,k=1}_n(\bm{r})$ reduce to the usual
spectroscopic amplitudes.  For a fixed $n$, on the other hand, the
$\phi^{A,k}_n(\bm{r})$ correspond to particle states built on
$|\Psi_{A-1}^n\rangle$.  Similarly, one can expand an ($A$-1)-body state,
$\Psi_{A-1}^n$, in terms of ($A$-2)-body basis states, $\Psi_{A-2}^m$, and
obtains expansion coefficients $\phi^{A-1,n}_m(\bm{r}) \equiv
\langle\Psi_{A-2}^{m}|a({\bm r}) | \Psi_{A-1}^n \rangle$. With respect to the
($A$-1)-body system, the ($A$-2)-body functions represent hole states and the
$A$-body functions are particle states.  The expansion coefficients
$\phi_m^{A-1,n}(\bm{r})$ can be used to rewrite the equations of motions
(eq.~(\ref{eq:schro})):
\bea
(E_A-E_{A-1}^n)\phi_n(\bm{r}) &=& - \frac{\nabla^2_{\bm{r}}}{2m} \phi_n(\bm{r})
 \nonumber\\&&
+ \sum_{n'=1}^\infty \sum_{m=1}^\infty \int d\bm{r}' \phi_m^{A-1,n*}(\bm{r}')
  \phi_m^{A-1,n'}(\bm{r}') V(|\bm{r}' - \bm{r}|) \phi_{n'}(\bm{r}) \; .
\label{eq:schro_s}
\eea                                
 
The kernel of the norm operator can also be written in terms of the particle 
or hole states.  For the particle states, we insert a complete set of 
intermediate $A$-body states, $\Psi_A^k(\bm{r}_1, \cdots,\bm{r}_A)$ in 
eq.~(\ref{eq:norm1}) to obtain:
\bea
{\cal N}(n,\bm{r},n',\bm{r}') &=& \sum_{k=1}^\infty
\phi_n^{A,k}(\bm{r})\phi_{n'}^{A,k*}(\bm{r}')   \; . 
\label{eq:ap1}
\eea
On the other hand, using eq.~(\ref{eq:bd}), the kernel of the norm operator 
can be expressed in terms of the (A-1)-body wave functions:
\bea
{\cal N}(n,\bm{r},n',\bm{r}') 
&=& \delta_{nn'}\delta(\bm{r}-\bm{r}')-  \nn\\
&& (A-1)\int \prod_{i=1}^{A-2} d\bm{r}_i 
\Psi_{A-1}^{n*}(\bm{r}_1,\ldots,\bm{r}_{A-2},{\bm r}')
\Psi_{A-1}^{n'}(\bm{r}_1,\ldots,\bm{r}_{A-2},{\bm r})  \; .
\label{eq:nam1}
\eea
This is not diagonal, in either $n$ or $\bm{r}$, even for the simplest systems.
From the last expression we derive the hole-state form:
\bea
{\cal N}(n,\bm{r},n',\bm{r}') &=&
\delta_{nn'}\delta(\bm{r}-\bm{r}')- \nn\\
&& \sum_{m=1}^\infty \phi_m^{A-1,n*}(\bm{r}')\phi_m^{A-1,n'}(\bm{r})\ .
\label{eq:a28}
\eea
Combining eqs.~(\ref{eq:ap1}) and (\ref{eq:a28}), we obtain a completeness
relation for the spectroscopic amplitudes corresponding to the set of particle
and hole states:
\bea
\delta_{nn'}\delta(\bm{r}-\bm{r}')=
\sum_{m=1}^\infty \phi_m^{A-1,n*}(\bm{r}')\phi_m^{A-1,n'}(\bm{r})
+\sum_{k=1}^\infty \phi_n^{A,k}(\bm{r})\phi_{n'}^{A,k*}(\bm{r}')  \; .
\label{eq:comp}
\eea
The sum over $m$ runs over all states of the $(A-2)$-body system [the hole
states of the $(A-1)$-body system] while the sum over $k$ runs over the states
of the $A$-body system [the particle states of the $(A-1)$-body system].  The
spectroscopic amplitudes for the particle states are not complete by themselves
since they lack the contributions that are Pauli blocked, namely those
contributions corresponding to hole states. Contrary to the impression that
this equation may give, the $\phi$'s are neither orthogonal nor normalized to
one.

To illustrate the formalism, we consider a two-particle system.  In this case,
the $\Psi_{A-1}^{n} (\bm{r}_1, \ldots,\bm{r}_{A-1}) \equiv
\Psi^n_1(\bm{r})$ are one-body wave functions and the kernel of the norm operator 
is given by: 
\bea
{\cal N}(n,\bm{r},n',\bm{r}') &=& \delta_{nn'}\delta(\bm{r}-\bm{r}') -
\Psi^{n*}_1(\bm{r}')\Psi^{n'}_1(\bm{r})  \; .
\eea
The terms diagonal in $n$ are projection operators onto states orthogonal to
$\Psi^n_1(\bm{r})$. This is also true for larger particle numbers if the
$\Psi_{A-1}^{n}(\bm{r}_1,\ldots,\bm{r}_{A-1})$ are single Slater determinants
[In that case ${\cal N}(n,\bm{r},n,\bm{r}')=\delta(\bm{r}-\bm{r}') -
\sum_{s=1}^A \phi^{n*}_s(\bm{r}') \phi^n_s(\bm{r})$, where the sum is over
occupied single-particle orbitals $\phi^n_s(\bm{r}')$].  While the matrix
elements diagonal in $n$ are projection operators by themselves, one needs to
divide the full norm operator by $A$ in order to obtain a projection operator.

\section{A bound for the spectroscopic factor}
\label{sec:App_norm}

In this appendix, we show that $S_n$ can be written as the product of two
factors, which express antisymmetry and dynamic distortion effects,
respectively.  When recoil and center-of-mass corrections are neglected, both
factors have to be less than or equal to one, yielding an upper limit of one
for the spectroscopic factor.

We start by defining a normalized spectroscopic amplitude $\tilde
\phi_n(\bm{r}) = \phi_n(\bm{r})/\sqrt{S_n}$ and express $\sqrt{S_n}$ as
follows:
\bea
\sqrt{S_n}&=&\int d\bm{r} \tilde \phi^*_n(\bm{r}) \phi_n(\bm{r}) 
   =\langle {\cal A}[ \tilde \phi_n \Psi_{A-1}^n] | \psi_A\rangle \; .
\eea
Next, we introduce a projection operator
\bea
P_n \equiv \frac{ |{\cal A}[ \tilde \phi_n \Psi_{A-1}^n]\rangle
  \langle {\cal A}[ \tilde \phi_n \Psi_{A-1}^n]| }{N_n} \;, 
\eea
where $N_n= \langle {\cal A}[ \tilde \phi_n \Psi_{A-1}^n] | {\cal A}[ \tilde
\phi_n \Psi_{A-1}^n]\rangle $.  An arbitrary state $|\psi_A\rangle$ can then be
broken into two orthogonal parts:
\bea
| \psi_A\rangle &=& P_n |\psi_A\rangle + (1-P_n) |\psi_A \rangle 
\equiv |\psi^{\parallel,n}_A\rangle + | \psi^{\perp,n}_A\rangle \; ,
\eea
where $|\psi^{\parallel,n}_A\rangle$ and $| \psi^{\perp,n}_A\rangle$
are the components of $| \psi_A\rangle$ which are parallel and 
orthogonal, respectively, to the state 
$|{\cal A}[ \tilde \phi_n \Psi_{A-1}^n]\rangle$.
We can then write the spectroscopic factor as:
\bea
S_n=N_n \langle\psi^{\parallel,n}_A |\psi^{\parallel,n}_A\rangle 
 =N_n \left( \langle\psi_A|\psi_A\rangle 
 - \langle \psi^{\perp,n}_A | \psi^{\perp,n}_A\rangle \right)  \; .
\label{eq:snorm}
\eea
The expression in brackets is less than or equal to one since
$\langle\psi_A|\psi_A\rangle =1$ and both $\langle \psi^{\parallel,n}_A |
\psi^{\parallel,n}_A \rangle $ and $\langle \psi^{\perp,n}_A | \psi^{\perp,n}_A
\rangle$ are positive semidefinite.  When the $(A-1)$-body system is
completely described by the wave function $\Psi_{A-1}^n$, i.e. when there are
no distortions due to the potential of the $A$-th nucleon, $\langle
\psi^{\perp,n}_A | \psi^{\perp,n}_A \rangle$ vanishes and $\langle
\psi^{\parallel,n}_A | \psi^{\parallel,n}_A \rangle =1$.  Thus $\langle
\psi^{\parallel,n}_A | \psi^{\parallel,n}_A \rangle$, and therefore $S_n$,
provides a measure of the dynamic distortions induced by the presence of the
extra particle.

Since the factor $N_n$ can be expressed as:
\bea
N_n = 1 - \sum_{m=1}^\infty \left( \int d\bm{r} \phi_m^{A-1,n*}(\bm{r}) 
\tilde \phi_n(\bm{r}) \right)^2  \;,
\label{eq:tnorm}
\eea
where the sum is explicitly non-negative and less than or equal to one, it is
also restricted, $N_n \leq 1$.  $N_n$ carries the effect of the
antisymmetrization and equals one only when $\phi_n(\bm{r})$ is orthogonal to $
\phi_m^{A-1,n}(\bm{r})$ for all $m$.

 From the above considerations it follows that $S_n \leq 1$.  If antisymmetry
was neglected, the spectroscopic factor could be as large as $A$, since the sum
rule given in eq.~(\ref{eq:pnorm}) would be the only restriction on $S_n$.
When center-of-mass corrections are incorporated, eq.~(\ref{eq:snorm}) still
holds, but eq.~(\ref{eq:tnorm}) has to be modified and $N_n$ can become larger
than one.  The influence of center-of-mass corrections is discussed in the next
Appendix.

\section{Center-of-mass corrections and intrinsic spectroscopic
amplitudes}
\label{App:COMcorr}

When dealing with the center-of-mass problem it is useful to introduce the
Jacobi coordinates $\rhov_j={\bm R}_j-{\bm r}_{j+1}$, where ${\bm R}_j$ is the
center-of-mass coordinate of the $j$-body system defined by particles $1$
through $j$. Taking into account the center-of-mass motion, the $A$ and
$(A-1)$-body wave functions are written as:
\bea
\psi_A({\bm r}_1,\cdots,{\bm r}_A)&=&
\frac{\exp[i {\bm k}_A \cdot {\bm R}_A]}{\sqrt {\cal V}}
\psi_A^I(\rhov_1,\cdots,\rhov_{A-1})
\eea
and
\bea 
\Psi^{n,k_{A-1}} (\bm{r_1},\cdots,{\bm r}_{A-1})
&=&\frac{\exp[i{\bm k}_{A-1} \cdot {\bm R_{A-1}}]}
{\sqrt {\cal V}} \Psi^{n}_I(\rhov_1,\cdots,\rhov_{A-2}) \; ,
\eea
respectively.  Here $k_A$ and $k_{A-1}$ are the center-of-mass momenta of the
$A$ and $(A-1)$-body systems, respectively. We have used box normalization with
volume $\cal V$. The spectroscopic amplitude is written as:
\bea
\phi_{n,k_{A-1}}(r)&=&\sqrt{A}\int \prod_{i=1}^A d{\bm r}_i\; 
 \delta({\bm r}-{\bm r}_A) \Psi^{n,k_{A-1}*}(\bm{r_1},\cdots,{\bm r}_{A-1})
\psi_A({\bm r}_1,\cdots,{\bm r}_A)\\
  &=& \frac{\exp[i{\bm r}\cdot({\bm k}_A - {\bm k}_{A-1})]}{\sqrt{{\cal V}}} 
\sqrt{A} \int \prod_{i=1}^{A-1} d{\bm \rho}_i\;
\frac{\exp\left[i \rhov_{A-1}\cdot \left(\frac{A-1}{A}{\bm k}_A-{\bm k}_{A-1}
\right)\right]}{\sqrt{\cal V}}\nonumber \\&&  \nn \\&&  
 \hspace{2cm}\times \Psi^{n*}_I(\bm{\rho_1},\cdots,{\bm
  \rho}_{A-2})\psi_A^I({\bm \rho}_1,\cdots,{\bm \rho}_{A-1}) \\ &=& \frac{
  \exp[i{\bm r}\cdot({\bm k}_A - {\bm k}_{A-1})]} {\sqrt{\cal
  V}}\tilde\phi_n^I\left(\frac{A-1}{A}{\bm k}_A-{\bm k}_{A-1}\right)  \; .
\label{eq:trans}
\eea
This equation is unexpected and requires some comments. Formally it is correct:
the spatial dependence of the spectroscopic amplitude is given by a plane wave
and the spectroscopic factor is $\left|\tilde\phi_n^I\left(\frac{A-1}{A}{\bm
k}_A - {\bm k}_{A-1}\right) \right|^2$. Since $\tilde\phi_n(k)$ is on the order
of $1/\sqrt{\cal V}$, it is small and the condition that the spectroscopic
factor must be less then or equal to one is easily satisfied.  The plane wave
behavior of the spectroscopic amplitude arises from translational
invariance. The combination $\left(\frac{A-1}{A}{\bm k}_A - {\bm
k}_{A-1}\right)$ is Galilean invariant.  By taking both the $(A-1)$-body and
the $A$-body systems to be in states of good momentum we have forced the $A$th
particle to also be in a state of good momentum; $\tilde\phi_n^I\left({\bm
k}\right)$ is then the probability amplitude for finding the $A$th particle
with relative momentum $\bm k$ when the $(A-1)$-body system is in state
$n$. Its Fourier transform, which we identify as the intrinsic spectroscopic
amplitude, is given by (compare eq.~(\ref{eq:b1b})):
\bea
\phi_n^I(\rhov)&=& \int d{\bm k}_{A-1} \frac{ \exp[i{\bm \rho}\cdot{\bm
k}_{A-1}]} {\sqrt{\cal V}}\tilde\phi_n^I\left({\bm k}_{A-1}\right) 
= \left. \int d{\bm k}_{A-1} \phi_{n,k_{A-1}}(\rhov) \right|_{k_A=0}
\\
&=&
\sqrt{A}\int \prod_{i=1}^{A-1} d{\bm
\rho}_i\;\delta(\rhov-\rhov_{A-1}) \Psi^{n*}_I(\bm{\rho_1},\cdots,{\bm
\rho}_{A-2})\psi_A^I({\bm \rho}_1,\cdots,{\bm \rho}_{A-1})
\eea
In analogy with $\tilde\phi_n^I\left({\bm k}\right)$, $\phi_n^I(\rhov)$ is the
probability amplitude for finding the $A$th particle at the distance ${\bm
\rho}$ from the center-of-mass of the $(A-1)$-body system when that system
is in the state $n$. The intrinsic $A$-body wave function can be written in
terms of the intrinsic spectroscopic amplitudes as (compare
eq.~(\ref{eq:expn})):
\bea
\psi_A^I(\rhov_1,\cdots,\rhov_{A-1})=\frac{1}{\sqrt{A}}\sum_{n=1}^\infty 
\phi_n^I(\rhov_{A-1})\Psi^{n}_I(\rhov_1,\cdots,\rhov_{A-2}).
\eea
As we show in the next paragraph, the intrinsic spectroscopic amplitude is
also the quantity that is needed to calculate physical observables.

We now write the transition matrix element, eq.~(\ref{eq:matrix}), in terms of
the intrinsic spectroscopic amplitude as follows:
\begin{eqnarray}
\langle \psi_A|\sum_{i=1}^A
      \exp[-i{\bm k} \cdot \bm{r}_i]|\psi^K_A\rangle 
 &=& \sum_{n,k_{A-1}}^\infty \int\;
      d\bm{r} \phi^*_{n,k_{A-1}}(\bm{r})
\exp[-i{\bm k} \cdot \bm{r}] \phi^K_{n,k_{A-1}}(\bm{r})\\
&=& \sum_{n=1}^\infty \int
      d\bm{\rho}\; \phi^{I*}_{n}(\bm{\rho})
\exp[-i{\bm k} \cdot \bm{\rho} (A-1)/A]
\phi^{KI}_{n}(\bm{\rho}) \nonumber \\&& 
\times \delta(k_K+k-k_A)  \; ,
\end{eqnarray}
where the transition operator has been taken to be a plane wave as is
appropriate for radiative capture and the spin and isospin dependencies have
been suppressed for simplicity. The functions $\phi^{KI}_{n} (\bm{\rho})$ and
$\phi^{I}_{n} (\bm{\rho})$ are intrinsic spectroscopic amplitudes for the
scattering and bound states, respectively.  The delta function ensures
overall momentum conservation. The $(A-1)/A$ factor in the exponential takes
care of the laboratory to center-of-mass transformation.

The equations of motion for the intrinsic spectroscopic amplitudes are easily
derived by substituting eq.~(\ref{eq:trans}) in eq.~(\ref{eq:schro_s}). 
This gives:
\begin{eqnarray}
(E_A^B-E_{A-1}^n)\phi_n^I(\bm{\rho}) &=& 
 - \frac{\nabla^2_{\bm{\rho}}}{2\mu} \phi_n^I(\bm{\rho}) 
 + \sum_{n'=1}^\infty \sum_{n'=1}^\infty \int d\bm{\rho}'
\rho_{nn'}({\bm \rho}') 
V(|\bm{\rho}' - \bm{\rho}|) \phi_{n'}^I(\bm{\rho})  \; ,
\label{eq:intschro}
\end{eqnarray}
where $\mu$ is the reduced mass and 
\bea
\rho_{nn'}({\bm \rho})=\left(\frac{A-1}{A-2}\right)^3
\phi_m^{I\;n*}(\bm{\rho} (A-1)/(A-2)) 
      \phi_m^{I\;n'}(\bm{\rho}(A-1)/(A-2)) 
\eea
is the transition density for the $(A-1)$-body system.  The $(A-1)/(A-2)$
factors originate in the conversion from the $\bm{R}_{A-2}-\bm{r}_{A-1}$ 
coordinate to the $\bm{R}_{A-1}-\bm{r}_{A-1}$ coordinate. 
The diagonal transition density is the usual density and is normalized to 
$A-1$.

The remaining quantity to consider is the norm operator, $\hat{{\cal N}}$.  
This is most easily done starting with eq.~(\ref{eq:a28}).  We obtain the 
following expression for the kernel ${\cal N}(n,{\bm r},n',{\bm r}')$:
\begin{eqnarray}
{\cal N}(n,\bm{r},n',\bm{r}') &=&
\delta_{nn'}\delta(\bm{r}-\bm{r}')- 
\left(\frac{(A-1)^2}{A(A-2)}\right)^3\nonumber\\&&\times
\sum_{m=1}^\infty
\tilde\phi_m^{n*}\left(\frac{(A-1)^2}{A(A-2)}\left(\bm{r}'+\frac{\bm
r}{(A-1)}\right)\right) \nonumber\\&&\hspace{1cm}\times
\tilde\phi_m^{n'}\left(\frac{(A-1)^2}{A(A-2)}\left(\bm{r}+
\frac{{\bm r}'}{(A-1)}\right)\right)
\end{eqnarray}
In contrast to the situation where the center-of-mass corrections are
neglected, the amplitudes in the sum given here depend on both coordinates.
Consequently, the intrinsic spectroscopic factors $S^I_n=\int d{\bm \rho}
\left|\phi_{n}^I(\bm{\rho})\right|^2$ no longer have to be less than one.  This
is illustrated in ref.~\cite{Die} for the harmonic oscillator model.  The
completeness relation for the particle and hole states, eq.~(\ref{eq:comp}), is
also modified, since this last equation must be used instead of
eq.~(\ref{eq:a28}). The spectroscopic amplitudes corresponding to the particle
states are just replaced by their intrinsic counterparts.

\end{document}